\documentclass[aps,pre,twocolumn]{revtex4-1}
\usepackage{amssymb}
\usepackage{amsmath,bm}
\usepackage{graphicx,color}
\usepackage{bbm}
\usepackage{multirow}
\newcommand{\C}[1]{{\mathcal{#1}}}

\newcommand{\im}{\text{Im}}

\newcommand{\ev}{\text{\bf E}}

\begin{document}

\title{The dimension of Diffusion Limited Aggregates grown on a line}
\author{Eviatar B. Procaccia$^{1,4}$ and Itamar Procaccia$^{2,3}$}
\affiliation{$^1$ Faculty of Industrial Engineering and Management, the Technion, Haifa 32000, Israel.\\$^2$Department of Chemical Physics, The Weizmann Institute of Science, Rehovot 76100, Israel.\\
$^3$ Center for OPTical IMagery Analysis and Learning, Northwestern Polytechnical University, Xi'an, 710072 China. \\
$^4$ Department of Mathematics, Texas A\&M University, College Station, Texas 77840, USA.}

\begin{abstract}
Diffusion Limited Aggregation (DLA) has served for forty years as a paradigmatic example for the creation of fractal growth patterns. In spite of thousands of references no exact result for the fractal dimension $D$ of DLA is known.  In this Letter we announce an exact result for off-lattice DLA grown on a line, $D=3/2$.
The result relies on representing DLA with iterated conformal maps, allowing one to prove self-affinity,  a proper scaling limit and a well defined fractal dimension. Mathematical proofs of the main results are available in \cite{20BPT}.
\end{abstract}

\maketitle

The diffusion limited aggregation (DLA) model was introduced in 1981 by Witten and Sander \cite{81WS}. The model has
been shown to underlie many pattern forming processes including dielectric breakdown \cite{84NPW}, two-fluid flow \cite{84Pat}, and
electrochemical deposition \cite{86GBCS}. The model begins with fixing
one particle at the center of coordinates in $d$ dimensions, and
follows the creation of a cluster by releasing a random walker 
from infinity, allowing it to walk around until it hits
any particle belonging to the cluster. Once there, the incoming particle is
attached to the growing cluster and a new one is released from infinity. The model was studied on
and off lattice in several dimensions $d \ge 2$; 
DLA has attracted enormous interest over the years since
it is a remarkable example of the spontaneous creation of
fractal objects. It is believed that asymptotically (when the
number of particles $N\to \infty$) the dimension D of the off-lattice cluster is
very close to 1.71 \cite{00DP,00DLP}, although there exists to date no rigorous proof for this fact. In addition, the model has attracted interest
since it was among the first \cite{86HMP} to offer a true multifractal
measure: the harmonic measure (which determines the probability that a random walker from infinity will hit a point at
the boundary) exhibits singularities that are usefully described using the multifractal formalism \cite{83HP,86HJKPS}. Nevertheless
DLA still poses more unsolved problems than answers. It is
obvious that a new language is needed in order to allow fresh
attempts to explain the growth patterns, the fractal dimension, and the multifractal properties of the harmonic measure. 
In this Letter we announce an exact result on the fractal dimension of DLA grown on a fiber. This model was simulated on the lattice
by Meakin in 1983 \cite{83Mea} with the numerical result that a typical tree with $N$ particles reaches a height (radius of gyration) of the order of $N^\delta$ with 
\begin{equation}
\delta \approx	0.665\pm 0.03 \ .
\end{equation}
Rewritten in terms of the fractal dimension of the clusters this translates to $D=1.50\pm 0.01$. Here we show that the DLA grown on a line off lattice has an exact dimension $D=3/2$.

The method used to establish this result is based on iterated conformal maps to grow a DLA cluster in a controlled fashion.
Introduced by Hastings and Levitov in \cite{98HL}, the idea is to employ a mapping $\phi_{\lambda,\theta}(\omega)$ that maps the exterior of the unit
circle to the exterior of unit circle with and added ``bump" or ``strike". This addition, whose linear size is $\lambda$, is placed
on the unit circle at a uniformly distributed  angle $\theta$. Iterating this mapping one defines a conformal
map $\Phi^{n}(\omega)$ according to
\begin{equation}
\Phi^{n}(\omega) \equiv \phi_{\lambda_1,\theta_1}\circ\phi_{\lambda_2,\theta_2}\circ \cdots \circ\phi_{\lambda_n,\theta_n} \ .
\label{defPhi}
\end{equation}
A major difficulty associated with the creation of the map for the classical example of DLA in two dimension has precluded so far the use of this method to determine exactly the fractal dimension of the growing cluster. The first difficulty is that the linear
size $\lambda$ has to be judiciously chosen in each step to grow a {\em fixed size} addition to the cluster,
\begin{equation}
\lambda_n = \frac{\lambda_0}{|\Phi^{{(n-1)}'}(e^{i\theta_n})|} \ .
\label{choicelam}
\end{equation}
Note the originally in Ref.~\cite{98HL} the size of $\lambda_n$ was allowed to vary, by taking the denominator in Eq.~(\ref{choicelam}) to the power
of $\alpha/2$. Thus the classical DLA model corresponds to $\alpha=2$.
A related difficulty lies in the monotonicity of the logarithmic capacity $c_n$. As $\omega\rightarrow\infty$, the Laurent expansion of $\Phi^n$ starts like
\begin{equation}
\Phi^{n}(\omega)=e^{c_n}\omega + O(1),
\end{equation}  
with $c_n>0$ and is monotonic increasing in $n$. In fact one can show that $c_n=(\log n)/D$ \cite{99DHOPSS,00DLP}. Thus one needs to normalize the
size of $\lambda$ more and more as the cluster grows. 

\begin{figure}
\includegraphics[width=0.45\textwidth]{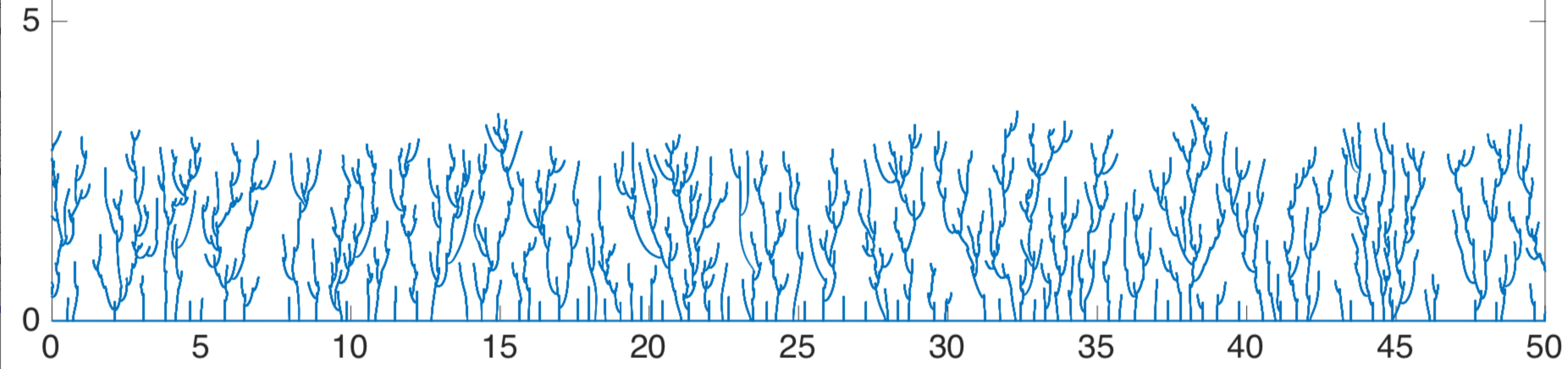}
\caption{Computer simulation of cluster growth in a window of the real axis}
	\label{fig:shl}
\end{figure}

Growing a DLA cluster on a fiber removes these difficulties altogether. One can map the upper half plane to the upper half plane
with a strike of size 1 above $x$ using the map $\phi_x(\omega)$
\begin{equation}
\phi_x(\omega)=x+\sqrt{(\omega-x)^2-1} \ .
\label{defphix}
\end{equation}
To represent the growth of a cluster on the a fiber one considers the whole upper half plane and orders the arrival times
of particles according to a homogeneous Poisson point process of intensity 1. Focus on a window of the real axis of length $R$
and mark the arrivals ot particles into positions $x_1, x_2,\cdots, x_k \cdots$ in this window, at times $0<t_1<t_2<\cdots<t_k \cdots$.   define
\begin{equation}
F_t^{(R)}(\omega)=\left\{\begin{array}{ll}
\omega & \text{ for }0\le t< t_1\\
\phi_{x_1}\circ \cdots \phi_{x_k}(z) & \text{ for }t_{k}\le t< t_{k+1}.
\end{array}\right.
\label{iterations}
\end{equation}
The wanted process is finally defined by taking the limit $F_t(\omega)=\lim_{R\to \infty}F_t^{(R)}(\omega)$. This process was proven to
exist in \cite{20BPT}, was denoted ``stationary Hasting-Levitov$(0)$" and shown to define a conformal map. In addition, this process is invariant to horizontal shifts. See Figure \ref{fig:shl} for a computer simulation of the process.

The relative simplicity of the resulting process is demonstrated by the Laurent expansion of $F_t(\omega)$. By
a tedious but straightforward calculation one shows that 
in the limit $\omega\rightarrow\infty$
\begin{equation}\label{laurent}
F_t(\omega)=\omega+\frac{i\pi}{2}t+O(1/\omega) \ .
\end{equation}
This is important since it implies that no repeated normalization of the strike sizes is necessary in this process.
Thus employing the power $\alpha=0$ is sufficient to generate a cluster growth in which the added particles
to the physical domain remain of fixed size. 
Moreover, it implies that redefining the map by inverting the order of iterations in Eq.~(\ref{iterations}) results,
in a fixed time $t$, in an inverted growth where the last particles grown normally appear first, and further particles
push them up in the half plane to end up with a cluster sharing the same distribution as the original one. 

Another immediate consequence of Eq.~(\ref{laurent}) is that the average height $h_c$ of the growth sites (known as the half-place capacity)
can be determined from the second term in the Laurent expansion \cite{14BN}:
\begin{equation}\label{height}
h_c=-i\lim_{\omega\rightarrow\infty}(F_t(\omega)-\omega)=\frac{\pi}{2}t \ .
\end{equation}
In other words, the positions in which random walkers coming from infinity meet the growing cluster increase
on the average linearly with time. 

Having control on the average height of the arrivals along the imaginary axis, we next focus on the fluctuations
of these arrivals along the real axis. To this aim we first consider the complex integral
\begin{equation}
\int_{-\infty}^\infty dx |\phi_x(\omega)-\omega|^2 \le \frac{C} {1+\im ~\omega} \le C\ ,
\end{equation}
where $C$ is a constant independent of $\omega$ and the last inequality stems from the fact that we work
in the upper half plane. Using this we can immediately derive 
a sharp estimate for the expectation of the fluctuations of the real part of the arrival points
$\ev (|F_t(\omega) -\omega-\frac{i\pi}{2}t|^2)$:
\begin{eqnarray}
&&\ev (|F_t(\omega) -\omega-\frac{i\pi}{2}t|^2)\equiv \\
&&\ev\Big(\int_0^t ds \int_{-\infty}^\infty dx ~|\phi_x(\omega)F_s(\omega) - F_s(\omega)|^2\Big) \le Ct \ . \nonumber
\label{nice}
\end{eqnarray}
We can deduce from Eq.~(\ref{laurent}) that the fluctuations in the real position of arrival of new particles grow like $\sqrt{t}$.
One should know however that this is an estimate of the global fluctuation over the whole real axis rather than on a single
growing tree. To achieve a statement about the fractal dimension requires a local result on the fluctuations. 

In order to find the number of particles added to a given tree in the cluster we consider the harmonic measure
of an interval on the real axis. Since particles are being added according to a homogeneous Poisson process, the harmonic measure
must be proportional to the length of the interval. Denote the harmonic measure of the interval $[a,b]$ at time $t$ as $\C H_{[a,b]}(t)$:
\begin{equation}
\C H_{[a,b]}(t) \equiv F_t^{-1}(b)-F_t^{-1}(a) \ .
\end{equation}
We will demonstrate now that for any chosen $a$ and $b$ the preimages $F_t^{-1}(a)$ and $F_t^{-1}(b)$ are  
diffusion processes for times of the order $t<(b-a)^2$. At time of the order of $(b-a)^2$ these diffusion
processes collide, at which point in time the harmonic measure $\C H_{[a,b]}(t)$ vanishes. Physically this means
that all the trees that grow out of the interval $[a,b]$ become shadowed by higher and broader trees and no new particle can ever reach
these trees. We note in passing that this result means that any set of trees that start to grow from any finite size interval
will eventually get shadowed and stop growing.  This is a warning that simulating on a fiber with periodic boundary conditions
is different, and will result in a single tree occupying all of the harmonic measure. It is remarkable that Meakin \cite{83Mea} had the intuition
to terminate his simulation at the ``right" time to get the correct result for this growth process!

The way that the Poisson process is defined it is clear that the number of particles arriving into any given area
in the upper half plane is proportional to that area. We know now that if we choose trees that start growing from an interval
$[a,b]$ of the order of unity, and condition on the harmonic measure not vanishing before or at time $t$, their typical height will be of the order of $t$. Moreover, tracing the area bounded between the curves defined by
$F_s^{-1}(a)$ and $F_s^{-1}(b)$ for $s\in [0,t]$, we know that this area scales like $t\times \sqrt{t}$,
and therefore the number of incoming particles belonging to the trees that survives until time $t$ is indeed
proportional to $t^{3/2}$. This provides the desired result that the height scales like $N^{2/3}$, or 
\begin{equation}
\boxed{\delta=2/3} \ .
\end{equation}
To discuss the dimension of the cluster we stress that the growing trees are not self-similar but rather self-affine. The Hausdorff
dimension therefore requires covering the set with different rescaling in the real and the imaginary directions. The scaling
is the natural one of a random walk, i.e. rescaling by $t$ in the imaginary direction by $\sqrt{t}$ in real direction. The result
then is that $D=3/2$. A rigorous proof of this result is Theorem 7.6 in \cite{20BPT}.

All the results presented above pertain to growth in all the upper half plane, and the relation to growth on a finite
fiber as executed in \cite{83Mea} must be discussed. Moreover, the simulations presented in \cite{83Mea} were done
{\em on lattice}  whereas the considerations above were all for random walks {\em off lattice}. Consider then
a cylinder of circumference of length $N$ (in units of the lattice constant) and infinite height. The process then involves
sending off $t\times N$ random walkers from infinity. To proceed we invoke the rigorous proof, cf. \cite{19MPZ,20PYZ}, that in the 
limit $N \to \infty$ the cluster that includes $t \times R$ particles grown over any finite interval of length $R\ll N$ is equivalent in all properties to a cluster grown over an interval of length $R$ belonging to the infinite real axis. Accordingly Meakin's simulation can be considered
relevant for DLA on-lattice growth on an infinite line. Since for the off-lattice growth we could show that trees
that contain $t$ particles are of height of the order of $t^{2/3}$ we now elaborate on Meakin's simulations and show an equivalent
result for the on-lattice simulation for large enough trees. 

\begin{figure}
	\includegraphics[width=0.50\textwidth]{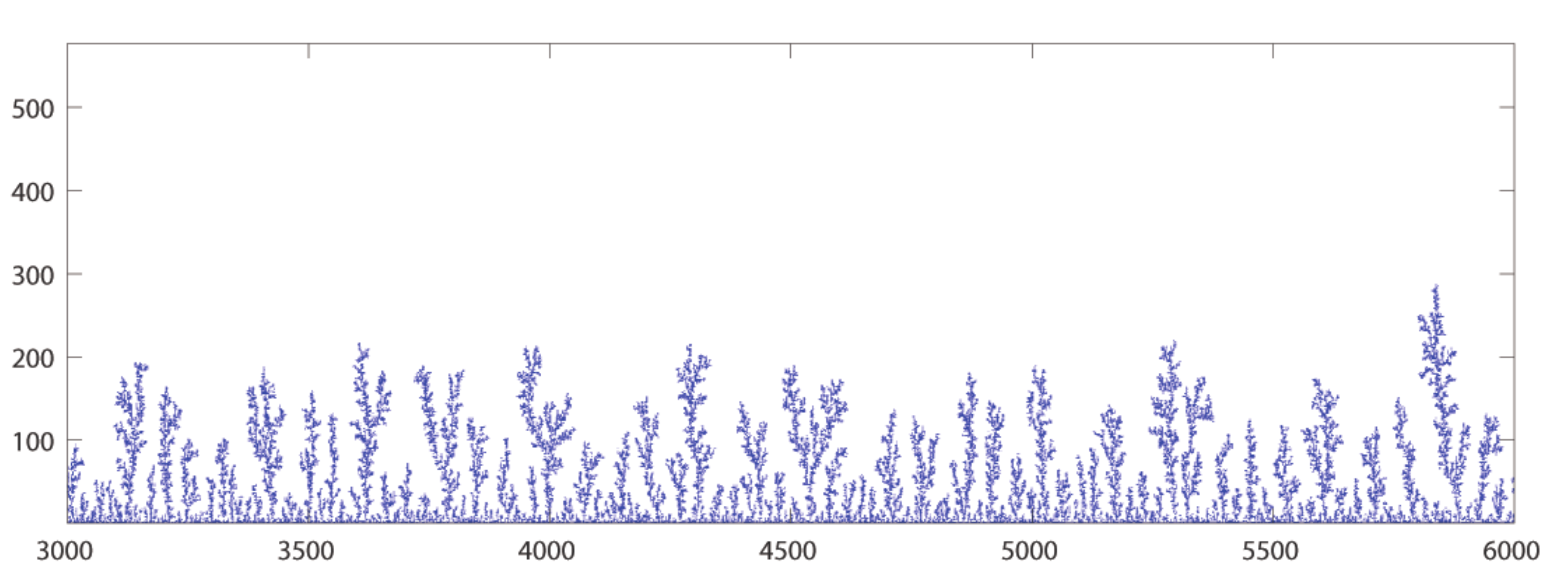}
	\caption{Computer simulation of cluster growth on lattice in a cylinder of circumference 8000 at $t=30$ (i.e. 240,000 particles).}
	\label{cylsim}
\end{figure}

The result of computer simulations on a cylinder of circumference 8000 at $t=30$ (i.e. 240,000 particles) is shown in Fig.~\ref{cylsim}.
The figure shows the cluster growth in the interval [3000,6000]. 
Contrary to Meakin who considered the radius of gyration of the whole cluster, we compute the height vs. the number of particles belonging
to individual trees. To accomplish this we identify each tree by the location of its root, paying attention to the particles added
to the same tree starting from this root. A log-log plot of the heights vs. the logarithm of the number of particles belonging to individual trees
\begin{figure}
	\hskip -1.0 cm
	\includegraphics[width=0.35\textwidth]{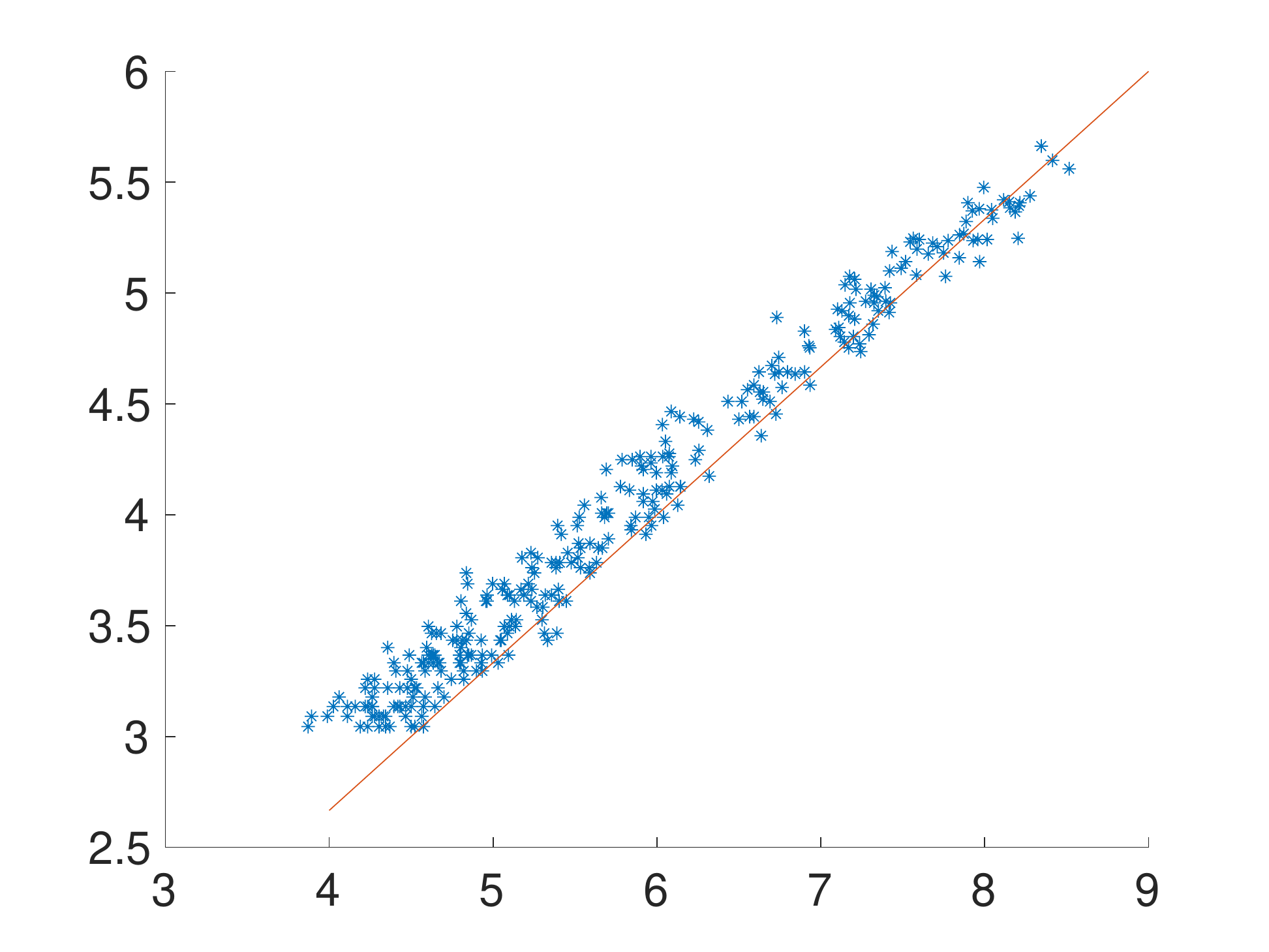}
	\caption{Measurement of the logarithm of the heights vs. the logarithm of the number of particles belonging to individual trees. The line
	is a guide to the eye showing the convergence to slope 2/3 for large trees.}
	\label{heightvsN}
\end{figure}
is shown in Fig.~\ref{heightvsN}. The expected slope of 2/3 is obtained asymptotically for large trees. Thus we can conclude from the
present simulation that the on-lattice model has the same Brownian fluctuations for the width of the growing trees. A tree will arrive
to a given height having a width that is determined by the distance between two Brownian paths conditioned on non-intersection. 

\begin{figure}
	\includegraphics[width=0.45\textwidth]{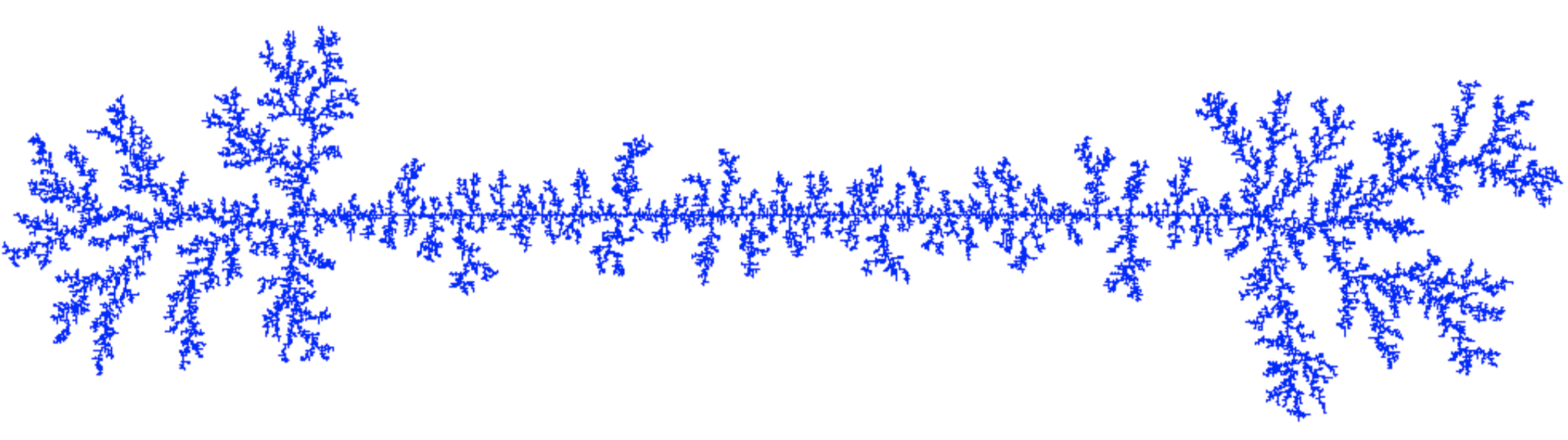}
	\caption{A typical cluster grown on a finite fiber embedded in two dimension. Only far away from the edges
the growing trees exhibit Brownian scaling. }
	\label{finite}
\end{figure}

Finally we should note that the pure Brownian scaling will fail in a finite cylinder when the simulation time gets too long.
When a given tree reaches the height of $\sqrt{N}$ then particles that might typically attach to this tree will already feel the
periodic boundary conditions. One expects that such a tree will occupy eventually the entire harmonic measure and all the other trees will not be
able to increase their width in subsequent times. Similar caution should be exercised for a growth on a finite sized
fiber (without boundary conditions). There the edges of the fiber will act as singular attracting points, and the growth far away
from the edges will exhibit Brownian scaling only for a finite time. An example of a simulation of growth on finite fiber is shown
in Fig.~\ref{finite}.

In summary, the DLA process over the real axis provides a relatively transparent example for the employment of iterated conformal maps
to represent the cluster growth. The reason for the relative ease is that the size of the strike does not depend on the
order of iteration, in contrast to the classical off-lattice DLA in two dimensions where the strike size changes in every iteration to conform
with the addition of a fixed size particle in the physical domain. As a consequence one can derive in the present case an exact result for the growth rate
and fractal dimension of the whole cluster or of individual trees. We note in passing that the dimension 3/2 was offered by Kesten as a rigorous
lower bound to the dimension of DLA grown on the square lattice in two dimensions \cite{87Kes,87Kesa}. It is known that DLA grown on the square lattice
in two dimensions looks asymptotically as a cross with four long arms \cite{89EMPZ,90EMPZ,17GB}. While the fractal dimension of the whole
cluster appears to exceed 3/2,  it is not impossible that further analysis might
lead to the possibility that the Brownian scaling is appropriate for individual trees growing far away from the tips, for reasons akin to the discussion offered above.

\bibliography{DLA}

\end{document}